\documentclass[12pt]{article}
\usepackage{graphicx,booktabs,amsmath,amssymb,multirow,natbib, adjustbox, xcolor,authblk}
\usepackage{psfrag, epsf}
\usepackage{pdfpages}

\newcommand{\blind}{0}

\definecolor{kitfarbe}{rgb}{0.004,0.588,0.51}
\definecolor{urlcol}{HTML}{294E85}
\definecolor{anccol}{HTML}{B52D26}  

\usepackage[colorlinks=true,
urlcolor=urlcol,
linkcolor=anccol,
citecolor=kitfarbe]{hyperref}       

\title{Learning to forecast: \\ The probabilistic time series forecasting challenge}

\date{\today}

\begin{document}
	
\def\spacingset#1{\renewcommand{\baselinestretch}%
	{#1}\small\normalsize} \spacingset{1}

\if0\blind
{
	\title{\bf Learning to forecast: \\ The probabilistic time series forecasting challenge}
	\author{Johannes Bracher$^{1,2}$, Nils Koster$^1$, Fabian Kr\"uger$^1$ and Sebastian Lerch$^{1,2}$\\
		$^1$Karlsruhe Institute of Technology\\ $^2$Heidelberg Institute for Theoretical Studies 
	}
	\maketitle
} \fi

\if1\blind
{
	\bigskip
	\bigskip
	\bigskip
	\begin{center}
		{\LARGE\bf Title}
	\end{center}
	\medskip
} \fi

\bigskip

\begin{abstract}
We report on a course project in which students submit weekly probabilistic forecasts of two weather variables and one financial variable. This real-time format allows students to engage in practical forecasting, which requires a diverse set of skills in data science and applied statistics. We describe the context and aims of the course, and discuss design parameters like the selection of target variables, the forecast submission process, the evaluation of forecast performance, and the feedback provided to students. Furthermore, we describe empirical properties of students' probabilistic forecasts, as well as some lessons learned on our part. 
\end{abstract}

\noindent%
{\it Keywords:} Statistics education, prediction, quantile modeling, time series analysis, model evaluation and selection
\vfill

\newpage

\section{Introduction}

Forecasting is a core task of statistics. Accordingly, the development, implementation and evaluation of prediction models plays a central role in statistics courses within degree programs such as economics, mathematics and computer science. Much of statistical forecasting takes place in a time series context, where the goal is to predict future observations of a univariate or multivariate time series. Reflecting the breadth and depth of time series analysis, there is a wealth of excellent textbooks, teaching materials, and university courses on the topic. For the sake of coherence and simplicity, however, even applied courses typically refer to idealized settings that do not match the requirements of real-world time series forecasting.

In particular, practical forecasting by definition refers to quantities not yet known at the time of prediction. Academic research and teaching, on the other hand, typically use historical data examples, and the quantities to be predicted are not actually unknown to the forecaster. The real-time setting can be mimicked, e.g., by considering a {rolling forecast origin}, sometimes also referred to as ``rolling window'' or ``rolling sample''. In this approach, the data are repeatedly split into training and test sets (using the $R$ newest observations up to time $t$ to predict the outcome at $t + 1$, observations up to $t + 1$ for the outcome at $t + 2$, and so forth). This, however, is only an imperfect imitation of the real-time setting. Firstly, modeling decisions are inevitably taken based on knowledge of the entire time series, thus including information which would not have been available in real time. Secondly, the procedure smooths over many imperfections of real-time data, which may be subject to initial errors, delays and revisions. (While some impressive efforts have been made to compile data sources that reconstruct forecasters' past information -- see, e.g., \citealt{Croushore2006} for economic examples -- the availability of such sources remains an exception.) Both aspects contribute to {hindsight bias}, i.e., after the fact the observed events may seem more predictable than they actually were. This tendency is further reinforced by the absence of time constraints in retrospective forecasting.

Practical forecasters need a diverse set of skills in statistics, involving theoretical knowledge (e.g., being able to select and test appropriate methodology), programming skills, as well as the ability to keep track of the forecasting workflow (ideally by using professional techniques like version control). Teaching these skills -- and practicing them ourselves -- was our main motivation for offering a course on real-time forecasting. As an additional feature, we sought to emphasize \textit{probabilistic} forecasts which acknowledge and quantify uncertainty \citep{GneitingKatzfuss2014}. These considerations led to the ``Probabilistic Time Series Forecasting Challenge'' that we describe in this paper. Its setup was inspired by collaborative forecasting efforts in epidemiology \citep[e.g.,][]{Cramer2022}. That said, several key modifications were necessary to make the project feasible and attractive for university students. In particular, in the hope of making the course as entertaining as possible, we introduced gamification-type aspects \citep{Gamification2,Gamification1} that are also present in data science competitions on the Kaggle platform \citep{Kaggle}.
We believe that the project-based character of the course promotes students' in-depth engagement with the statistical techniques they employ (see, e.g., \citealt{Cetinkaya2021} and \citealt{Raschka2022} for similar assessments). In particular, the process of producing and validating specific numbers to be submitted as forecasts requires students to delve into the details of their forecasting method. This type of engagement, which we consider essential in applied statistics, is easily brushed over in more traditional course formats. All rules and procedures for the challenge were pre-defined in a public preregistration which was available to participants \citep{preregistration}.

This paper describes our experiences and learnings from running this novel course format at the Karlsruhe Institute of Technology (KIT) in Germany between October 2021 and February 2022. (The second edition of the course took place in very similar form between October 2022 and February 2023.) The paper contributes to the statistics literature in two main ways. Firstly, while several collaborative forecasting projects contain real-time and/or probabilistic elements (see Section \ref{sec:comparison}), our project translates the real-time probabilistic forecasting setup into a university course. While we are aware of real-time forecasting projects run by colleagues as one element of a course, these projects were simpler and not publicly documented. We thus hope that our experiences and materials are of interest to instructors who consider offering a similar course at their institution. As illustrated by our interdisciplinary choice of target variables from two distinct fields (weather and finance), we expect that similar course offerings are worthwhile in many application areas. Secondly, while probabilistic forecasts have recently become more popular, much of applied statistics is still dominated by point forecasts. Collecting probabilistic forecasts is hence a non-trivial task, not only in educational settings but also in the context of professional forecasters. In economics, for example, professionals' point forecasts have clearly been more successful than their probabilistic forecasts \citep{FaustWright2013, Krueger2017, Clements2018, GlasHartmann2022}. Furthermore, epidemiological expert forecasts have repeatedly been found to be strongly overconfident \citep[e.g.,][]{Bracher2021,Cramer2022}. Several design parameters we cover (such as the choice of forecast evaluation criteria and feedback mechanisms) are relevant beyond the educational context. Our in-depth discussion may hence hence provide guidance for designing collaborative forecasting projects with professional participants, too. 

The paper is organized as follows. Section \ref{sec:struct} describes the structure of the course, thereby highlighting relevant design parameters. Section \ref{sec:comparison} compares the course format to existing collaborative forecasting projects. Section \ref{sec:empirical} presents empirical results on participants' forecasts, and Section \ref{sec:disc} provides a concluding discussion. 
To avoid clutter, we collect web resources in Table \ref{tab:urls} of the appendix, and refer to them as ``URL1'' to ``URL15'' in the text. Additional empirical analyses can be found in the Online Supplement. Forecast and outcome data are available on the course's GitLab repository at \hyperref[tab:urls]{URL1}. Code to replicate the empirical analysis in the present paper is available at \hyperref[tab:urls]{URL2}. An interactive visualization of the forecasts and outcomes is available at \hyperref[tab:urls]{URL3}.

\section{Structure of the forecasting challenge}\label{sec:struct}

In this section, we describe the context of the course, as well as its collaborative forecasting process. 

\subsection{Context}

\paragraph{Goal of the course.}

The goal of the course was to let students explore the process of making and evaluating forecasts in real time. By the end of the course, students should have developed a practical and yet principled forecasting workflow, using statistical tools and strategies of their choice. In particular, this involves developing practically viable solutions under time constraints, as well as critically assessing and improving their methodology in an iterative fashion. Furthermore, students should be able to clearly describe and motivate their approach in a seminar paper (see Section \ref{subsec:submit_eval}). As detailed throughout Section \ref{sec:struct}, we sought to enable students to reach this learning goal by providing suitable background materials, a fun yet functional setup for forecast submission and evaluation, as well as regular opportunities to discuss among each other and with instructors.

\paragraph{Background of participants.} 
Bachelors and Masters students enrolled at our institution in the degrees of industrial engineering and management, computer science and business mathematics were eligible to apply for the course as part of an elective module. In the Bachelors program, these degrees contain at least three semester-long courses on each of mathematics and computer science, and two semester-long courses on statistics (e.g., descriptive statistics, basic probability theory, hypothesis testing and linear regression). As a prerequisite for participation, we further required ``proficiency in a programming language like R, Python, or Matlab'', which students can acquire in various courses at our institution. Among 21 applicants, we admitted two Bachelors, twelve Masters, and four PhD students (with the latter not required to pass an examination). Of these 18 students, 17 participated continuously throughout the challenge. Admissions were based on knowledge of statistics (as documented by a transcript of records submitted by applicants), as well as experience in programming and applied statistics (as stated by applicants in a short letter of motivation). Many participants had practiced statistics or machine learning beyond the introductory courses mentioned above, e.g., in other elective courses or their Bachelors thesis. The high acceptance ratio of 18/21 can be explained by the fact that the course is part of a methodological elective module within the different degree programs, resulting in a fairly specialized and highly motivated target audience.

\paragraph{Provided material.} 
In preparation, we recorded several video lectures for the students. Students could hence re-watch the lectures, a substantial benefit as we anticipated a steep learning curve in the first few weeks of the course. As detailed in Appendix \ref{sec:appendix_schedule}, the lectures contained basics in regression modeling, probabilistic forecasting, time series analysis and real-time forecasting as well as pointers to R \citep{R} and Python \citep{Python} resources. 
While we did not restrict the students' choice of programming language, our support was limited to R and Python.
We further provided code for the respective benchmark models (all three variables), for advanced benchmark models (for temperature and wind, see below), as well as some data loading scripts and scripts to check the correct format of a submission file. 

\subsection{Targets and benchmarks}
\label{sec:targets}
\begin{figure}[h]
    \centering
    \includegraphics[width = 0.95\textwidth]{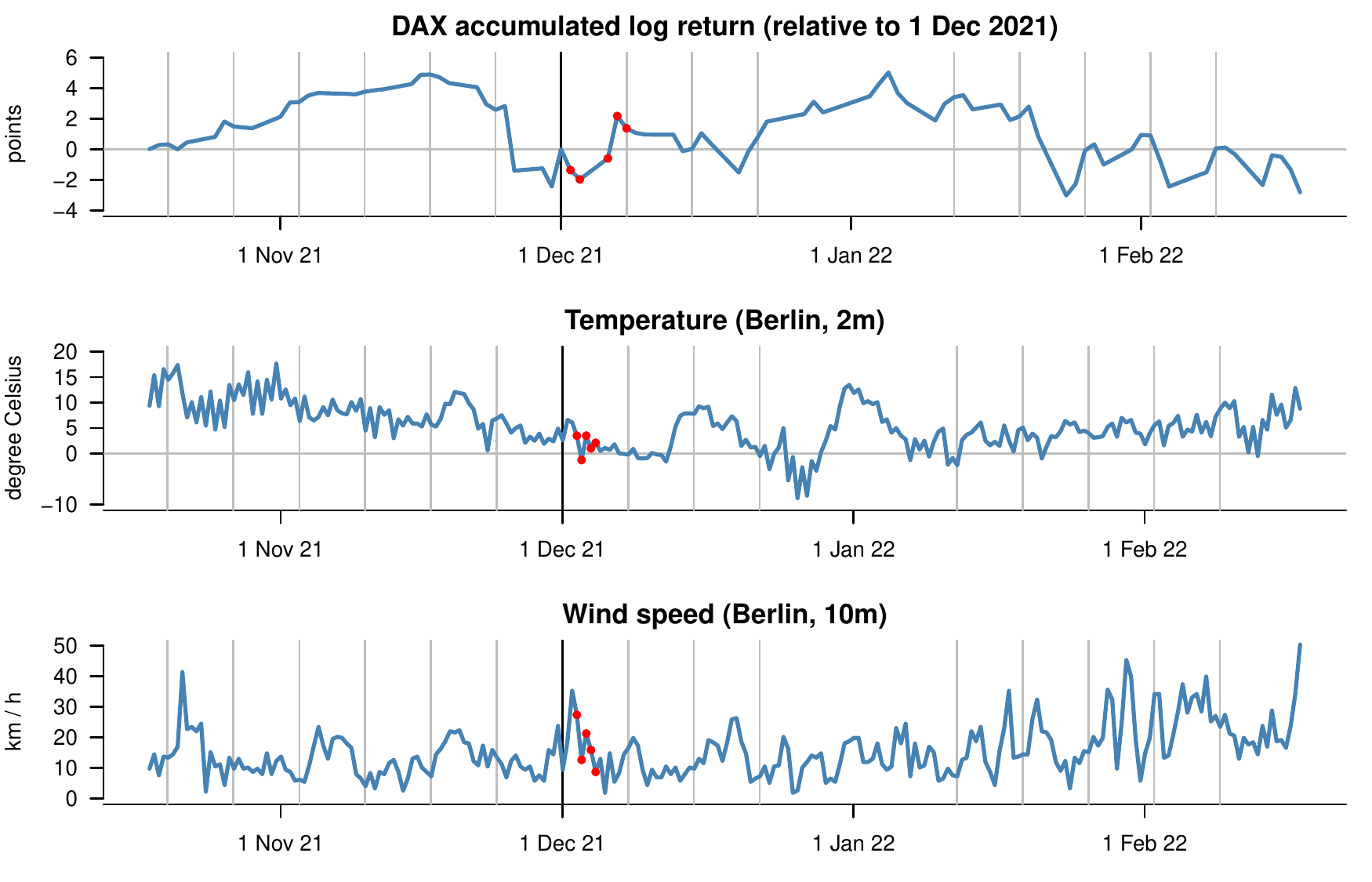}
    \caption{Time series to be predicted: accumulated log returns of the DAX stock index, temperature 2 meters above the ground and wind speed 10 meters above the ground in Berlin (note: only values at 0:00 and 12:00 UTC shown). Grey vertical lines show time points at which forecasts were made. To illustrate the defined forecast lead times, the black vertical lines show 1 December and the red dots the quantities which needed to be predicted on 1 December.}
    \label{fig:data}
\end{figure}

\paragraph{Target Variables.} Each Wednesday, participants were required to provide forecasts for the following three target variables (see Section \ref{subsec:submit_eval} for details on the submission process):
\begin{itemize}
    \item accumulated \textit{log-return} of the German stock index (DAX) of the next five business days (defined as 100 times the difference between the log-transformed value of the DAX at the target time and its last known value at the time of prediction). We used Yahoo Finance (\hyperref[tab:urls]{URL4}) as an easily accessible data source.
    \item \textit{temperature} 2 meters above ground, measured in °C, for a weather station close to Karlsruhe (Rheinstetten) for 12:00 UTC on the respective upcoming Thursday, Friday and Saturday as well as 0:00 UTC on the respective upcoming Thursday and Friday.
    \item \textit{wind speed} 10 meters above ground, measured in km/h, for the same weather station and horizons as for temperature.
\end{itemize}
Figure \ref{fig:data} shows the time series of the three targets and illustrates the chosen lead times. As the goal of the challenge was probabilistic forecasting, we required forecasts for the respective 2.5\%, 25\%, 50\%, 75\% and 97.5\% quantiles for said targets. Participants were allowed to use any additional data sets available to them. 

By choosing these targets, we aimed for a diverse set of forecasting tasks in a classical time series format. The data are relatively easy to access and understand, and constructing plausible statistical forecasts is feasible for university students.

Unfortunately, the Rheinstetten weather station, which had been chosen for its geographic proximity to our institution, did not report data due to a connectivity problem during the first week of the challenge. We therefore switched to the more distant Berlin-Tempelhof station (\hyperref[tab:urls]{URL5}) whose observations displayed broadly similar empirical properties.

\paragraph{Setup for numerical weather predictions (NWP).} 
Nowadays, weather forecasts are typically based on numerical weather prediction (NWP) models which represent physical processes via systems of partial differential equations. Weather services usually run an ensemble of NWP model simulations, e.g.,\ based on randomly perturbed initial conditions. To reflect the common practice in weather forecasting for the temperature and wind speed targets in the setting of the challenge, we utilized NWP ensemble forecasts from the operational weather prediction model of the German weather service (Deutscher Wetterdienst, DWD), the ICON (ICOsahedral Nonhydrostatic) model \citep{ICON}. 
The ICON model is run at a 20 km resolution over Europe and contains 40 ensemble members. 
While ICON forecast data is openly available via DWD's {Climate Data Center} (\hyperref[tab:urls]{URL6}), downloading and handling these large datasets (e.g., to extract forecasts at specific locations) can be cumbersome and requires specialist knowledge. 
Using automated downloading and pre-processing scripts run on a remote server, we provided up-to-date ICON ensemble forecasts to all participants. The forecasts were disseminated on a daily basis in the form of text files via a dedicated internal GitLab repository, and contained predictions of seven weather variables (mean sea level pressure, total cloud cover, direct downward shortwave radiation, temperature at 2 meters and at 850 hPa, wind speed and maximum wind gust at 10 meters) with lead times up to 120 hours for the target location(s). 

NWP ensembles are typically biased and lack calibration. They thus require correction via so-called post-processing methods. The distributional regression approaches employed in operational weather forecasting and research range from simple statistical techniques to advanced machine learning methods, see \citet{VannitsemEtAl2021} for a recent overview. 
One of the most popular approaches is the ensemble model output statistics (EMOS) method proposed by \citet{GneitingEtAl2005} where a parametric distribution is assumed for the target variable. The distribution parameters are then modeled as functions of summary statistics from the NWP ensemble predictions. 
To enable the implementation of post-processing models for the weather targets, we compiled an additional dataset of past ICON ensemble predictions and corresponding observations at the target locations from December 2018 to September 2021 which was provided to the participants, along with implementations of standard EMOS models for temperature and wind based on the crch software package \citep{crch} for R.

\paragraph{Benchmark models.} For each target variable, we defined a benchmark model in the study preregistration. These benchmark models, which are summarized in Table \ref{tab:bench}, were intended to put participants' forecast performance into perspective (e.g., by calculating skill scores). Our benchmark models were chosen to be simple and easy to understand, while still being empirically plausible. In addition to the benchmark models, we generated EMOS forecasts for the two weather targets. All benchmark forecasts were based on code that we also provided to participants. 

\begin{table}[h]
\centering
\caption{Summary of benchmark models.}
\label{tab:bench}
\begin{tabular}{ll}
\toprule
Variable & Description of benchmark \\ 
\midrule
DAX & Empirical return quantiles, computed from a rolling \\
& sample $\mathcal{S}$ of $1\,000$ days, i.e., $\mathcal{S} = \{r_{t-999,h}, \ldots, r_{t,h}\},$ \\
& with $r_{t,h} = 100 \times (\log \text{DAX}_t - \log \text{DAX}_{t-h})$\\[.1cm]
Wind & Empirical quantiles of raw ensemble forecast \\[.1cm]
Temperature & Empirical quantiles of raw ensemble forecast \\\bottomrule
\end{tabular}

\end{table}

\paragraph{Ensemble forecasts.} 
Ensembles (or ``combinations'') are a common and empirically successful tool to summarize a set of forecasts of the same target variable \citep[e.g.,][]{GneitingRaftery2005,timmermann2006forecast,WangEtAl2022}. In our context, ensembles summarize the numerous contributed predictions and illustrate the potential of collaborative forecasting. Specifically, we generated two different forecast ensembles:
\begin{enumerate}
    \item \textit{Mean ensemble}: The $\alpha$ quantile of the ensemble is given by the mean of the respective $\alpha$ quantiles of its members.
    \item \textit{Median ensemble}: The $\alpha$ quantile of the ensemble is given by the median of the respective $\alpha$ quantiles of its members.
\end{enumerate}

\cite{LichtendahlEtAl2013} show that averaging quantile forecasts (as in our mean ensemble) yields more concentrated forecast distributions than ``linear'' combination methods that average the cumulative distribution functions (CDFs) associated with individual forecasts. This feature is advantageous if the individual forecast distributions are calibrated, but can also be disadvantageous if the individual forecast distributions are overconfident. In the context of our course project, quantile-based ensembles are much more practical than linear combinations since individual forecasters submit quantiles at specific levels but not full CDFs. While we are not aware of theoretical results on using the median (rather than the mean) of individual quantile forecasts, the latter approach is more robust to outliers in the individual forecasts; see \citet{Bracher2021} for empirical illustrations in the context of COVID-19.\\ 

As specified in our preregistration of the course project, we initially planned to consider an inverse score-weighted ensemble in addition. However, we ultimately refrained from adding this layer of complexity as we found the simpler mean and median ensembles well sufficient for the purpose of summarizing collaborative forecast performance in the course. Our use of equal weights is also in line with a large literature that documents the typically good empirical performance of equal weights as compared to more sophisticated weighting schemes (e.g., \citealt[Section 2.2]{WangEtAl2022} and \citealt[Section 1]{ZischkeEtAl2022}).

\subsection{Submission and evaluation} \label{subsec:submit_eval}

\paragraph{Submission} The project took place throughout the whole winter semester 2021/2022, where the first submission had to be handed in by October 27th, 2021. 
Due to the initial weather station's connectivity problems noted above, weather forecasts started only at November 3rd, 2021. The last submission deadline was February 9th, 2022. Excluding a two-week Christmas break, this resulted in 14 weeks of DAX forecasts and 13 weeks of weather variable forecasts. 
Each Wednesday, forecasts needed to be sent to us via email in a specified .csv-format by 11:59 PM. Table \ref{tab:example_submission} shows an illustrative submission file for November 3rd, 2021.
The latest available ensemble weather forecasts were those from the NWP model runs initialized on Wednesday at 00 UTC. The first weather targets to be predicted corresponded to observations made on Thursday at 12 UTC, which is 36 hours after the NWP model's initialization time and 12 hours after the submission deadline.
The participants' forecast files were then uploaded Thursday morning on our public repository. Students were allowed to skip submissions twice without failing the course.

In order to protect the students' anonymity, each student was asked to pick a character from five fictional universes as their alias: 
{Discworld} (\hyperref[tab:urls]{URL7}, not chosen at all), 
{Brooklyn Nine-Nine} (\hyperref[tab:urls]{URL8}, chosen three times), 
{Friends \& Joey} (\hyperref[tab:urls]{URL9}, chosen five times), 
{Game of Thrones} (\hyperref[tab:urls]{URL10}, chosen five times) 
and {Star Wars} (\hyperref[tab:urls]{URL11}, chosen seven times). 

Each Thursday, course participants and one or two instructors met online to discuss the past week of forecasting. This allowed students to learn from each other's experiences and obtain ideas to improve their forecasting approach. Although the call was not mandatory, participation remained high during the whole semester. Furthermore, we provided an online course forum that was used actively throughout the semester.

\begin{table}
\centering
\caption{Example of a correctly specified submission file for November 3rd, 2021.}
\label{tab:example_submission}
\begin{adjustbox}{max width=\textwidth}
\begin{tabular}{lllccccc}
\toprule
\textbf{forecast\_date} & \textbf{target} & \textbf{horizon} & \textbf{q0.025} & \textbf{q0.25} & \textbf{q0.5} & \textbf{q0.75} & \textbf{q0.975} \\ \midrule
2021-11-03 & DAX & 1 day & -1.8 & -0.3 & 0.1 & 0.6 & 1.7\\
2021-11-03 & DAX & 2 day & -3.0 & -0.5 & 0.2 & 0.9 & 2.0\\
2021-11-03 & DAX & 5 day & -3.0 & -0.7 & 0.2 & 1.2 & 2.4\\
2021-11-03 & DAX & 6 day & -3.6 & -0.9 & 0.3 & 1.2 & 2.7\\
2021-11-03 & DAX & 7 day & -3.6 & -0.9 & 0.5 & 1.4 & 3.2\\
\addlinespace
2021-11-03 & temperature & 36 hour & 6.5 & 8.0 & 8.6 & 9.2 & 10.4\\
2021-11-03 & temperature & 48 hour & 6.2 & 7.9 & 8.7 & 9.2 & 10.6\\
2021-11-03 & temperature & 60 hour & 7.9 & 9.8 & 10.9 & 11.7 & 13.4\\
2021-11-03 & temperature & 72 hour & 4.3 & 6.8 & 7.6 & 8.3 & 9.7\\
2021-11-03 & temperature & 84 hour & 8.5 & 10.4 & 11.3 & 12.0 & 14.2\\
\addlinespace
2021-11-03 & wind & 36 hour & 8.7 & 13.8 & 16.5 & 19.4 & 26.2\\
2021-11-03 & wind & 48 hour & 5.8 & 15.5 & 18.9 & 23.1 & 30.8\\
2021-11-03 & wind & 60 hour & 9.7 & 14.2 & 16.7 & 19.0 & 23.8\\
2021-11-03 & wind & 72 hour & 6.9 & 11.9 & 14.2 & 17.1 & 24.3\\
2021-11-03 & wind & 84 hour & 8.9 & 14.4 & 17.7 & 20.8 & 26.3\\ \bottomrule
\end{tabular}
\end{adjustbox}
\end{table}

\paragraph{Public display and evaluation of forecasts} An {interactive visualization tool} (\hyperref[tab:urls]{URL3}) implemented in R Shiny \citep{shiny} and plotly \citep{plotly} was made available and updated after each round of submissions. This allowed students to browse their own and others' past forecasts and visually compare them against later observed data. The website also included a regularly updated leader board covering the various employed evaluation measures. The app was hosted using the shinyapps.io service, with the underlying code and data available at \hyperref[tab:urls]{URL12}. 

\paragraph{Performance measures.} Several different statistical measures were used to assess the probabilistic forecasts submitted by the participants. These were specified prior to the start of the course in the study preregistration \citep{preregistration}. The primary outcome measure was the linear quantile score
$$
\text{QS}_\alpha(q_\alpha, y) = 2 \times \{\mathbf{1}(y < q_\alpha) - \alpha\} \times (q_\alpha - y),
$$
where $q_\alpha$ is the submitted predictive $\alpha$ quantile with $\alpha \in \{0.025, 0.25, 0.5, 0.75, 0.975\}$ and $y$ is the observed value. The linear quantile score is a proper scoring rule \citep{Gneiting2007}, meaning that it encourages ``honest'' forecasting and is immune to hedging. For each forecast target this score was averaged across the five submitted quantile levels, which results in an approximation of the widely-used continuous ranked probability score \citep{Laio2007}. In addition, the absolute error and coverage percentages of the central 50\% and 95\% prediction intervals were assessed (see \cite{Gneiting2023} for a more detailed overview of evaluation methods for quantile forecasts).

In order to summarize forecast performance, it seems natural to aggregate the scores across targets and forecast horizons. However, simple averages of scores are hard to interpret since the targets (and the corresponding quantile scores) are measured in different units. Furthermore, averaging likely introduces undesired weighting effects as the level and variability of scores may differ considerably across targets and horizons. In order to avoid these drawbacks, we used a rank-based approach. Specifically, for each target type $t$ (i.e., DAX, temperature and wind speed) and forecast horizon $h$ we first computed average scores
$$
\bar{S}_{t, h} = \frac{1}{N_t}\sum_{i = 1}^{N_t} S_{t, h, i}
$$
across the $N_t$ submission weeks, where $N_t = 14$ for DAX and $N_t = 13$ for the weather variables. To simplify interpretation and to improve comparability across targets and horizons, we then translated these scores into skill scores,
\begin{equation}
S^*_{t, h} = 1 - \frac{\bar{S}_{t, h}}{\bar{S}^\text{bench}_{t, h}}    
\label{eq:skill_score}
\end{equation}
where $\bar{S}^\text{bench}_{t, h}$ is the average score achieved by the respective benchmark model. For each target and horizon, participants were then ranked by the skill score (or, equivalently, the average score). To obtain an overall ranking, we averaged each participant's ranks for the three targets and five horizons. In case of equal average ranks, the best achieved rank was used as a tiebreaker, followed by the average rank achieved for temperature and ultimately a coin flip. 

We note that from a theoretical perspective, incentives in forecast competitions are complex and depend on participants' preferences (e.g., risk attitudes) and the details of the forecasting setup (e.g., dependent versus independent outcomes across submission rounds). While the recent literature has made intriguing progress \citep[e.g.,][]{PfeiferEtAl2014,FrongilloEtAl2021,WitkowskiEtAl2022}, it does not yet provide clear guidance for complex practical settings like ours. We thus chose the pragmatic methodology described above. In Section \ref{sec:ranking} and the Online Supplement, we provide a comparison to other ranking methodologies. 

If participants did not submit forecasts for one or more targets, their score for each target was imputed with $1.01$ times the worst average linear quantile score achieved by any participant for that target. This system was chosen to incentivize regular participation.

\paragraph{Examination and grading.} Grading was based on a seminar paper (due at the end of the course) in which participants summarized their prediction approach, technical pipeline and learnings from the challenge. Students were further required to submit the program code corresponding to their final submission in the challenge. Forecast performance by itself did not matter for grading. However, the process of forecasting and updating (as described in the seminar paper) should be plausible and convincing. Finally, there were small awards for the best-performing participants.

\begin{figure}
    \centering
    \includegraphics[width = 0.95\textwidth]{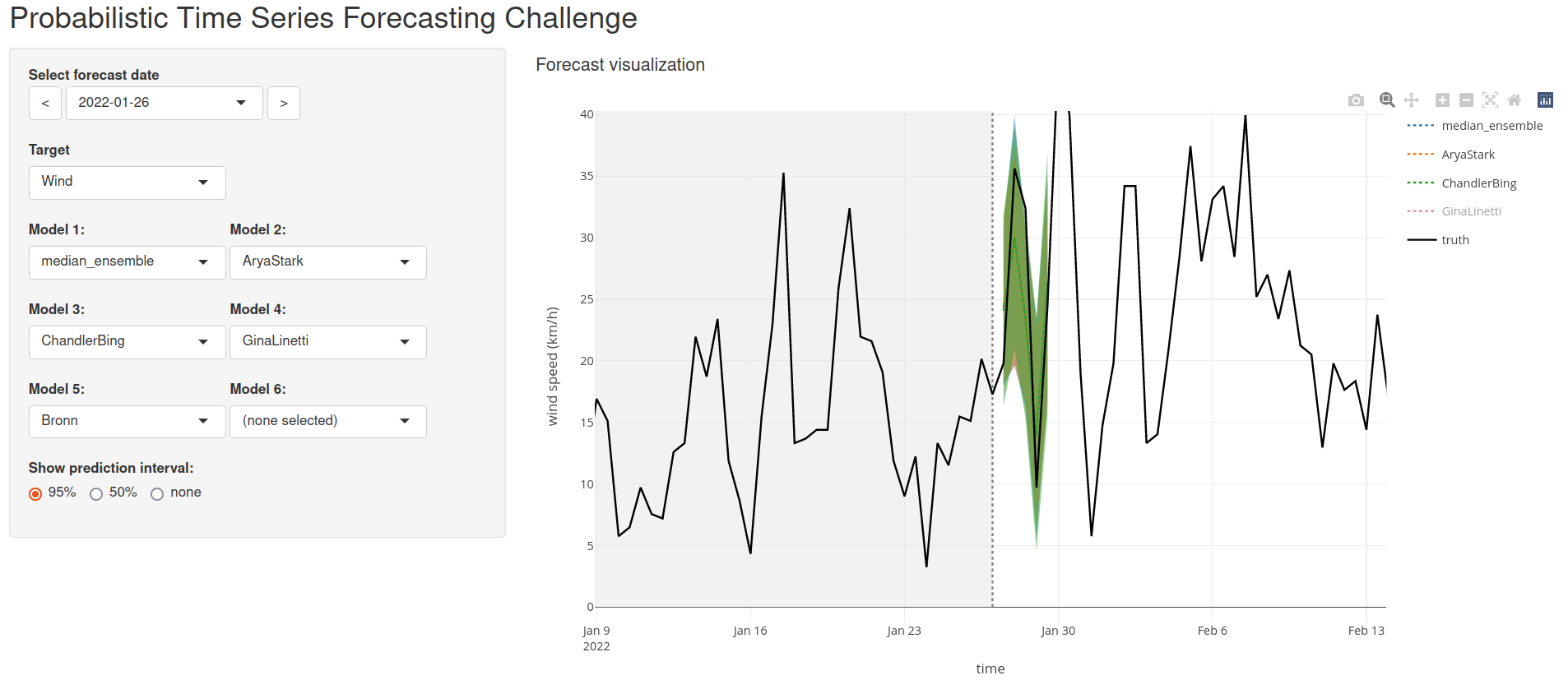}
    \caption{Interactive graphic display showing wind speed forecasts made by participants and the later observed time series.}
    \label{fig:screenshot_website}
\end{figure}

\section{Comparison to collaborative forecasting projects}\label{sec:comparison}

We next relate our setup to collaborative forecasting projects across different areas of applied statistics. While these projects have been framed in somewhat different ways (as ``competitions'', ``surveys'', and ``hubs''), a key commonality among them is that several participants make forecasts of the same target variable. 

\subsection{Forecasting competitions}

Forecasting competitions have found widespread use in statistics, complementing large-scale benchmarking and model comparison efforts in machine learning (such as the {ImageNet Large Scale Visual Recognition Challenge} at \hyperref[tab:urls]{URL13}) and other fields. Forecasting competitions have the potential to improve the theory and practice of forecasting, for example by motivating novel methodological developments. For recent overviews, we refer to \citet{Hyndman2020}, \citet[Section 2.12.7]{PetropoulosEtAl2022} and \citet{MakridakisEtAl2022}. In addition to forecasting competitions organized in an academic context, there is a plethora of competitions available on platforms such as Kaggle \citep{Kaggle}. Most forecasting competitions focus on deterministic rather than probabilistic forecasts, and few use a real-time format or focus on time series aspects. We discuss three notable exceptions in the following.

\paragraph{Global Energy Forecasting Competitions.} The Global Energy Forecasting Competition (GEFCom), a series of three international competitions received considerable interest in energy forecasting research. The focus of GEFCom2014 \citep{GEFCom} was on probabilistic forecasts in the form of quantiles in four tracks on load, price, wind, and solar forecasting. The competition used historic data, but was conducted in a pseudo real-time setting, with weekly submissions of forecasts based on incremental data releases. 

\paragraph{Competitions on subseasonal weather forecasting.} While subseasonal weather forecasts with forecast horizons of around 2 to 6 weeks are of considerable importance for many application areas such as agriculture, energy or health, NWP models typically lose most of their predictive ability on those time-scales \citep{WhiteEtAl2022}. Recent international competitions on subseasonal weather forecasting have thus focused on improving those predictions, in particular through the use of machine learning methods. 
The Subseasonal Climate Forecast Rodeo (S2S Rodeo) was a real-time forecasting competition organized by the U.S.\ Bureau of Reclamation and the National Oceanic and Atmospheric Administration that took place over a year from April 2017 to April 2018 \citep{S2SRodeo}. Participants had to submit deterministic forecasts of average temperature and accumulated precipitation over the western contiguous U.S.\ 3--4 and 5--6 weeks ahead. 

In a conceptually similar setup, the World Meteorological Organization (WMO) organized the {WMO Prize Challenge to Improve Sub-Seasonal to Seasonal Predictions Using Artificial Intelligence} (\hyperref[tab:urls]{URL14})  in 2021 \citep{WMOS2SAI}. Probabilistic predictions of categorical events (below normal / normal / above normal) had to be submitted for the same target variables for all land grid points on the entire globe. Unlike the S2S Rodeo, the competition was not conducted in a real-time setting. 
Within both competitions, datasets of historic and current ensemble forecasts from NWP models were provided to the participants.

\paragraph{M6 Financial Forecasting Competition.} The recent M6 competition \citep{MOFC2022a,MOFC2022b} took place from February 2022 to January 2023 in a real-time format, with twelve monthly rounds of forecast submissions. It is the most recent edition of a series of forecasting competitions with varying goals and data \citep[see, e.g.,][]{Hyndman2020,MakridakisEtAl2022}. Submissions to the M6 competition include probabilistic forecasts of stocks' relative performance (e.g., the probability that the given stock's return is among the top quintile of all assets considered), as well as a hypothetical investment decision (the portfolio weight assigned to each asset). Correspondingly, the competition's prizes reward both statistical forecasting performance (as measured by the Ranked Probability Score due to \citealt{Epstein1969}) and investment performance (as measured by the risk-return relationship of the chosen portfolio). 

\subsection{Survey of Professional Forecasters (SPF)} Initiated already in 1968, the SPF is a quarterly forecast survey covering US macroeconomic outcomes like the gross domestic product or the inflation rate \citep{Croushore2019}. A European analogue of the survey was established by the European Central Bank in 1999 \citep{BowlesEtAl2007}. The SPF's strict real-time schedule and its coverage of probabilistic forecasts are similar to our course. On the other hand, many macroeconomic variables are measured only quarterly and are published with a delay of several weeks. In consequence, obtaining feedback on forecast performance is much harder for SPF participants than for forecasters in fields like meteorology, where outcome data are available at high frequency. 

\subsection{COVID-19 Forecast Hubs} The overall format of the challenge as well as parts of the technical infrastructure were inspired by ongoing forecasting activities on the COVID-19 pandemic. Specifically, one member of the organizing team was involved in the operation of various \textit{Forecast Hub} platforms \citep{Cramer2022, Bracher2021, Sherratt2022}, which collate and combine short-term forecasts of COVID-19 case, hospitalization and death numbers. These platforms operate using a similar quantile-based format with weekly submissions and multiple lead times. 

\section{Empirical results}\label{sec:empirical}

In this section, we describe empirical properties of students' probabilistic forecasts.

\subsection{Forecasting approaches used by participants}
In the following, we present the forecasting methods undertaken by the participants throughout the challenge. 
Table \ref{tab:forecasting_appr} summarizes the implemented approaches, as indicated in the students' final project reports (excluding PhD students, who did not have to submit reports). We group approaches into their broader methodological classes, that is, we do not distinguish between different parameter or architecture choices. The row on ``additional data'' indicates whether students considered feature variables constructed from external data sources, which was allowed by the rules of the course.
\begin{table}[h]
\centering
\caption{Forecasting approaches implemented by the students grouped into broad categories for each target variable. The first number denotes how often the method was implemented, while the second number denotes how often the method performed best in the students' final evaluation.}
\label{tab:forecasting_appr}
\begin{tabular}{lccc}
\toprule
                            & Wind & Temperature  & DAX \\ \midrule
Baseline                    & 3/2  & 3/2          & 4/3 \\
EMOS                        & 8/6  & 8/7          &  -  \\
Quantile Regression         & 2/1  & 2/1          & 9/2 \\
Quantile Random Forest      & 4/0  & 3/0          & 2/0 \\
Gradient Boosting           & 4/2  & 5/2          & 4/2 \\
Neural Network              & 3/1  & 4/1          & 4/3 \\
AR(I)MA                     & 1/0  & 1/0          & 4/1 \\
GARCH                       & 0/0  & 0/0          & 5/3 \\
Other                       & 2/2  & 2/2          & 3/2 \\
Ensemble of Models          & 2/2  & 1/1          & 3/3 \\
Additional Data             & 0/0  & 0/0          & 4/2 \\ \bottomrule
\end{tabular}

\end{table}

Students treated the wind and temperature targets similarly. 
For those targets, EMOS is by far the most frequently implemented and best final model. Interestingly, for two students the raw ensemble performed best. We suspect that this is due to programming errors, as EMOS did indeed outperform the raw ensemble (see Section \ref{subsec:forecast_perf} for details). Other methods, such as quantile regression and quantile random forests were tried a few times, but apparently did not perform as well. Model classes that do not fit the weather forecasting task, specifically AR(I)MA and GARCH were rarely used. No participant tried to find additional datasets, which seems plausible given the rich information set condensed by the ensemble forecasts. Instead of considering additional data, students tried to gain performance by adapting the model (class). Some students did not only utilize the ensemble forecasts of the respective target variable, but also included other variables (e.g., cloud cover) in their predictions.

For the DAX, where no feature data was provided, students' approaches were more heterogeneous than for wind and temperature. No model stands out as a clear favorite. While most students implemented  quantile regression methods, few chose them as their final model. Besides the  (possibly adapted) benchmark model, neural networks, GARCH and ensembles of models performed best for the students. Four students used additional data sets.

Regarding their challenges during the semester, students mostly named technical problems with software packages and implementation errors in their own code. Given that students were required to develop their own code, this was to be expected as the workload remained high during the semester. This is a key difference to most other challenges (e.g., on Kaggle) where the real-time aspect is missing. Lastly, some of the implementation choices and method descriptions in the reports suggest conceptual misunderstandings on the part of participants. Similar to occasional coding errors, this finding is not surprising given the difficulty and volume of the course content. 

\subsection{Evaluation sample}

Our forecast evaluation sample comprises 14 weeks for the DAX (starting with the submissions on October 27, 2021) and 13 weeks for the weather targets (starting with the submissions on November 3, 2021). For all targets, the final submission round was on February 9, 2022, and submissions were paused during a Christmas break. The later submission start for the weather targets was due to the switch to the Berlin weather station described in Section \ref{sec:targets}. This setup yields a total of 199 forecast/observation pairs (DAX: 14 weeks times 5 horizons, minus one banking holiday on Christmas; two weather targets: 13 weeks times five horizons). 

\subsection{Forecast performance}
\label{subsec:forecast_perf}
Table \ref{tab:coverage} presents the coverage rates of the benchmark (see Table \ref{tab:bench} for a summary) and ensemble forecasts. Coverage is defined as the share of test-sample observations that are within the prediction interval. For example, the coverage rate for the 50\% interval should be 50\%. For brevity, we focus on the mean ensemble (see end of Section \ref{sec:targets}). The results for the median ensemble are qualitatively similar. Since each individual coverage rate shown in Table \ref{tab:coverage} is computed from a rather small sample of $13$ or $14$ weeks, we limit ourselves to a broad summary interpretation. The DAX benchmark generally has satisfactory coverage close to its nominal level. This result is unsurprising, given that the benchmark is an estimate of the unconditional return distribution. While the latter distribution is not sharp (in particular, it uses no conditioning information), it is known to satisfy various notions of forecast calibration, including interval coverage that we consider here \citep[see, e.g.,][]{GneitingKatzfuss2014}. For temperature, the benchmark is overconfident at the shortest horizon (with coverage rates falling short of their nominal levels), but achieves satisfactory coverage rates otherwise. For wind, the benchmark seems overconfident at all horizons. For the ensemble forecasts, Table \ref{tab:coverage} shows no clear signs of miscalibration, with coverage rates being reasonably close to their nominal levels for all targets and horizons.

\begin{table}[htbp]
\centering
\caption{Coverage rates of prediction intervals for each respective horizon, in percent. ``Ensemble'' denotes mean ensemble, ``Temp'' denotes temperature.  \label{tab:coverage}}
\begin{tabular}{llrrrr}
\toprule
&& \multicolumn{2}{c}{Coverage ($50 \%$ level) [\%]} & \multicolumn{2}{c}{Coverage ($95 \%$ level) [\%]}  \medskip \\
Target & Horizon & Benchmark & Ensemble & Benchmark & Ensemble \\
\midrule
DAX & 1 day & 71.4 & 71.4 & 100.0 & 100.0\\
DAX & 2 day & 50.0 & 42.9 & 92.9 & 92.9\\
DAX & 5 day & 50.0 & 50.0 & 92.9 & 85.7\\
DAX & 6 day & 35.7 & 35.7 & 100.0 & 85.7\\
DAX & 7 day & 53.8 & 46.2 & 100.0 & 100.0\\
\addlinespace
Temp & 36 hour & 15.4 & 30.8 & 61.5 & 92.3\\
Temp & 48 hour & 46.2 & 76.9 & 92.3 & 100.0\\
Temp & 60 hour & 53.8 & 76.9 & 100.0 & 100.0\\
Temp & 72 hour & 30.8 & 53.8 & 76.9 & 92.3\\
Temp & 84 hour & 69.2 & 61.5 & 92.3 & 92.3\\
\addlinespace
Wind & 36 hour & 7.7 & 61.5 & 53.8 & 100.0\\
Wind & 48 hour & 23.1 & 46.2 & 46.2 & 92.3\\
Wind & 60 hour & 30.8 & 61.5 & 61.5 & 84.6\\
Wind & 72 hour & 30.8 & 38.5 & 53.8 & 92.3\\
Wind & 84 hour & 53.8 & 30.8 & 61.5 & 92.3\\
\bottomrule
\end{tabular}

\end{table}

\begin{figure}[htbp]
    \centering
    \includegraphics[width = \textwidth]{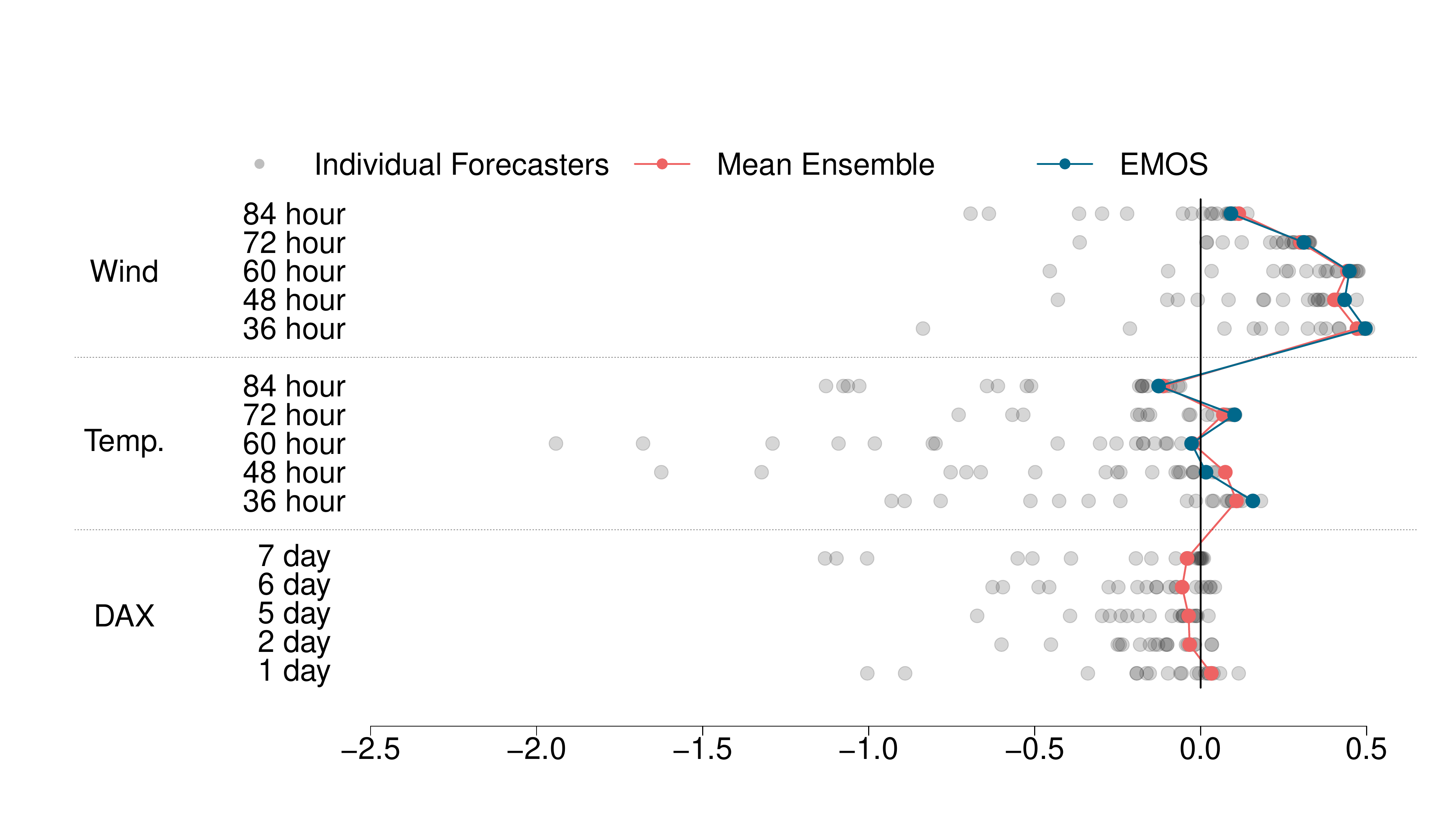}\vspace{-.4cm}
    \caption{Graphical summary of forecast performance. Dots are skill scores relative to benchmark model. Larger skill scores are better and a skill of 0 corresponds to the performance of the benchmark.\label{fig:skill_scores}}
\end{figure}

Figure \ref{fig:skill_scores} summarizes performance as measured by the quantile scoring function, based on which we compute skill scores relative to the benchmark forecasts (see Equation \ref{eq:skill_score}). A positive skill score indicates that a forecast outperforms the benchmark. For DAX, a majority of forecasters performs worse than the benchmark. Furthermore, the mean ensemble performs similar to the benchmark. For wind, we observe the opposite result: Many individual forecasters, as well as the mean ensemble, outperform the benchmark. The results for temperature are between these two polar cases: Whereas many individual forecasters (and the ensemble) outperform the benchmark of horizons of 36, 48 and 72 hours, the benchmark performs very well at the other two horizons of 60 and 84 hours. For both weather targets, the mean ensemble performs similar to the EMOS model. For all three targets, the mean ensemble performs similar to the best individual forecasters, and is remarkably robust to the inclusion of forecasters who perform below average. Similar ``wisdom of the crowds'' type results on forecast combination have been documented for various application areas and types of forecasts \citep[see][Section 2.6.4, for a brief review]{PetropoulosEtAl2022}.

The results in Table \ref{tab:coverage} and Figure \ref{fig:skill_scores} are closely in line with each other, in the following sense: for targets and horizons at which the benchmark lacks calibration (i.e., wind at all and temperature at some horizons), we observe that ensemble forecasts improve upon the benchmark by restoring calibration. Conversely, the benchmark seems hard to beat for the remaining targets and horizons. This observation emphasizes the role of calibration for making good predictions. In principle, it is also possible to improve upon the benchmark by making sharper predictions, subject to remaining calibrated \citep[see][]{GneitingKatzfuss2014,Pohle2020,KruegerZiegel2021}. Such an improvement could occur by using a larger information set than the benchmark, e.g., by making appropriate use of additional regressor variables. In order to investigate this possible channel, Table 1 of the Online Supplement considers the length of the benchmark's and ensemble's prediction intervals. For the weather targets,  the ensemble's intervals are generally wider than the benchmark's intervals. For the DAX, the ensemble's intervals are slightly shorter, but the ensemble does not outperform the benchmark. These results suggest that the sharpness channel just described plays a limited role in our setting, at least for the mean ensemble. 

\subsection{Exploratory analysis of learning effects}

The sequential (time series) forecasting process is characteristic of our course, and distinguishes it from ``static'' prediction contests where participants are assigned a fixed training and test sample at the same time. In principle, this sequential process makes it possible to learn about the success or failure of alternative forecasting methods as time proceeds. One can hence think of two sources of data that participants can use for model validation: Historical data (available at the beginning of the course) and real-time data (arising as the course proceeds). Ex ante, we found it hard to judge the practical relevance of real-time data, especially in the light of the rather short sample period of 14 weeks and possible heterogeneity across forecast targets and horizons. 

\begin{figure}[h]
    \centering
    \includegraphics[width=.7\textwidth]{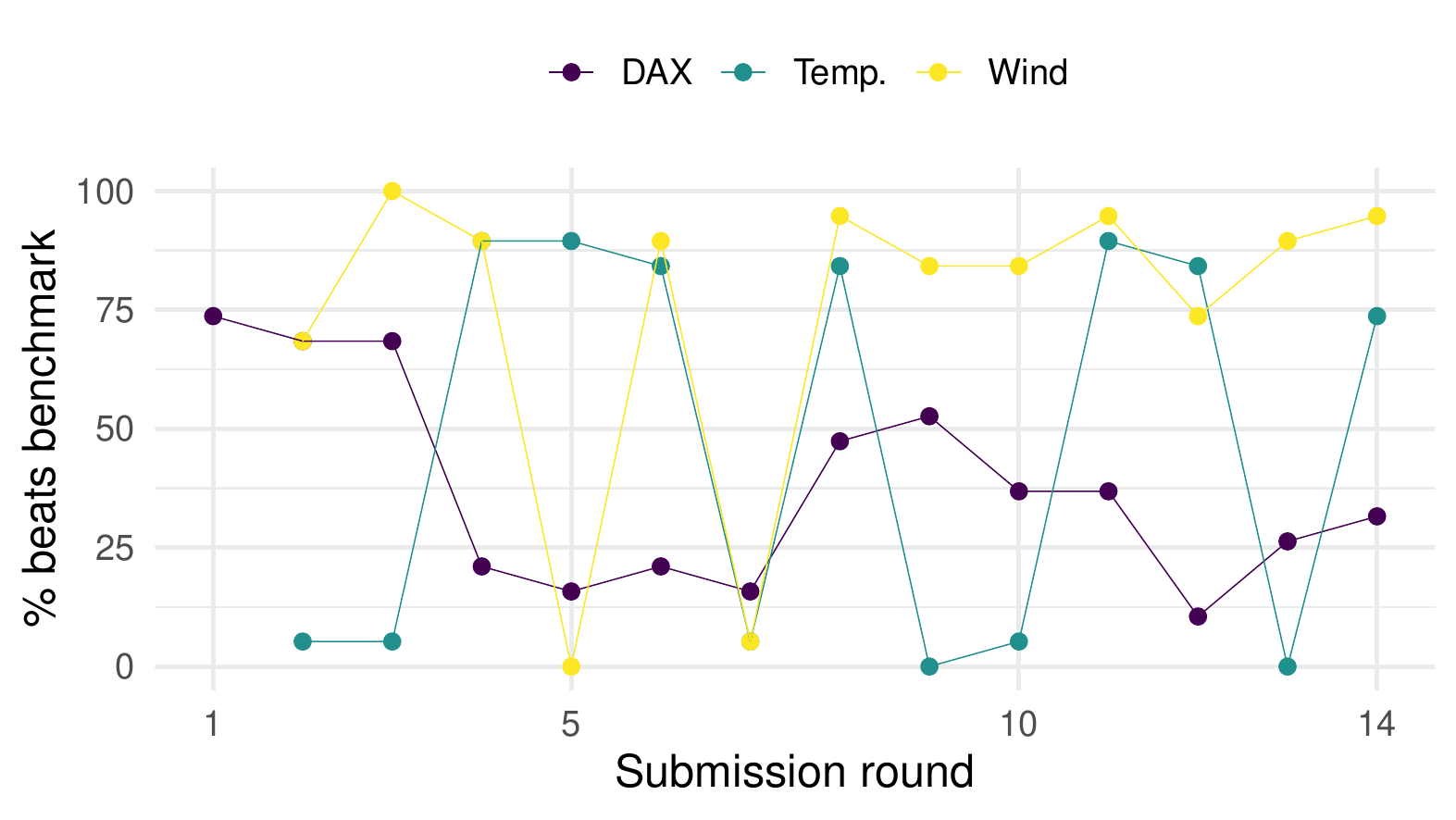}
    \caption{Share of forecasters who beats the benchmark, for each target and submission round.}
    \label{fig:share_beats_benchmark}
\end{figure}

In order to explore the relevance of real-time learning, we compute the share of participants who outperform the benchmark in each week of the course. We label this analysis, as well as the the one in Section \ref{sec:ranking}, as ``exploratory'' since it was not specified in our preregistration plan but was devised during the semester. For a given week and forecast target (suppressed for ease of notation), we compute the following share:
$$\frac{1}{n}\sum_{i=1}^n  \mathbf{1}\left\{\frac{1}{5}\sum_{h=1}^5 S_{h}^\text{bench} > \frac{1}{5}\sum_{h=1}^5 S_{h,i}\right\}$$
where $n$ denotes the number of participants for that week. In words, the indicator function refers to the event that forecaster $i$ attains a smaller (i.e., better) score than the benchmark, on average across the five forecast horizons. Under the assumption that real-time data enable learning over time, we would expect the share in question to increase over time, in that forecasters learn how to beat the benchmark. Figure \ref{fig:share_beats_benchmark} indicates that this phenomenon of an increasing share is not observed for either of the three targets. In light of these results, we conjecture that real-time learning effects are limited in our setup, and may be dominated by other effects, such as participants' use of more complex methods as time proceeds.

\subsection{Exploratory analysis of ranking methodology}\label{sec:ranking}

Our overall ranking of forecast performance intended to provide simple-to-interpret feedback to participants, in addition to more granular information on quantile scores for the different target variables and forecast horizons. The ranking also contributes to the course's gamification aspect. Since there are several plausible methods for constructing such a ranking, Section 2 of the Online Supplement investigates various methods and assess their sensitivity. From a practical perspective, the empirical results are similar across the methods considered. In particular, summing participants' skill scores (see Equation \ref{eq:skill_score}) across all targets and horizons yields very similar results as the rank-based methodology that we used in the course's web app described in Section \ref{subsec:submit_eval}. 

\section{Discussion}\label{sec:disc}

We conclude by discussing student reactions as well as possible adaptations of the course format. 

\subsection{Student reactions}

\paragraph{Student feedback.} 
In the following, we briefly summarize students' feedback in an anoynymous mid-semester course evaluation survey. All students stated that that the amount of work for the course was large or very large, and a clear majority of students stated that they learned a lot during the course. In their textual comments, a majority of students expressed that they liked the free nature, practical relevance, and steep learning curve of the course, and appreciated the support of the team. Several students highlighted the value of exchanging and discussing strategies among each other during weekly calls.
Conversely, some students did not like the steep learning curve at the beginning of the course, found the amount of information in the video lectures overwhelming, or suggested that the course requirements were not communicated clearly enough. Presumably, our stated prerequisites of a ``good working knowledge in statistics'' and ``proficiency in a programming language'' should have been more specific indeed. For example, instructors might use example quiz questions or programming tasks in order to illustrate the background knowledge that they expect from participants. In order to ease participants' workload at the beginning of the course, instructors may consider making materials (such as video lectures) available early so that participants can study them well before the first round of forecast submissions. 

\paragraph{Students' choice of forecasting method.} 
As evidenced by their final reports, students' motives for choosing a specific forecasting method can roughly be grouped into three categories: Firstly, empirical performance in cross-validation experiments, or in predicting new data that became available over time. Secondly, accounting for domain-specific data structures (such as modeling the conditional variance of financial returns). Thirdly, curiosity to experiment with a specific approach or model class. In each case, students' thinking was likely shaped by the background material provided in the video lectures, as well as joint discussions in the weekly online meetings and the course forum. 

Many approaches used by participants seem overly complex in hindsight, likely indicating a tendency towards overfitting and a lack of rigorous comparison of the devised methods to the provided benchmarks. By contrast, simple forecasting methods (careful extensions of the benchmarks) performed best. Apart from avoiding overfitting, simpler approaches also make it easier to spot and fix implementation errors. Indeed, some students found it challenging to develop program code for their preferred forecasting method, or to translate existing code from another programming language into their preferred language. 

Potential reasons for participants' preference for more complex models include prior exposure to machine learning techniques, usually in idealized and/or big data settings where those tend to work well. Furthermore, curiosity and gaming aspects made it natural to move away from the simple benchmarks. Finally, the hesitancy to use simple models might be due to a fear of not using ``state of the art'' approaches and not having done ``enough'' to attain a good grade.

From the perspective of an individual forecaster who wants to optimize their performance (in terms of the quantile score, say), the use of simple models would have been advisable in retrospect. From a broader perspective, the implications are more subtle. Firstly, for an individual participant, experimenting with complex approaches may yield higher learning gains, and be more entertaining, than selecting simple and conservative methods from the onset. Secondly, from the perspective of a meta-forecaster who seeks to optimize ensemble performance, a diverse set of (possibly uncalibrated) forecasts may well be preferable to a homogeneous set of calibrated forecasts. Indeed, Figure \ref{fig:skill_scores} indicates that the mean ensemble outperforms the vast majority of individual participants, and compares favorably to the benchmarks in the case of the weather variables. These results are much in line with ``wisdom of the crowd'' type effects discussed earlier. 

In response to the above, we plan to clearly communicate the following points in future course offerings. Firstly, we suggest that students incorporate simple plausibility checks (such as visual comparisons to the benchmarks) into their forecasting workflow. Secondly, we emphasize that students should not feel pressured to use complex methods for the sake of using complex methods. Finally, we encourage students to choose among the models implemented in their preferred programming language (mostly R or Python), as translating code or even switching languages can be time consuming and error prone in the present context. Apart from this advice, we think that the heterogeneity in students' motives and forecasting approaches is desirable, and should not be restricted by overly specific instructions. 

\subsection{Possible adaptations of the course format}

\paragraph{Substantive context of target variables.} The choice of target variables is highly flexible and can be adapted depending on participants' interests and backgrounds. While the course described here focused on meteorology and finance, we have replaced the temperature target by an hourly energy demand time series in the course's second offering (taking place from October 2022 through February 2023), and have found it to be equally suitable. Similarly, we expect that variables from epidemiology (e.g., case numbers of a disease), business (e.g., counts of product sales) and other fields are worth considering. See Section \ref{sec:targets} for further discussion and practical considerations regarding the choice of target variables. 

\paragraph{Schedule of submissions.} The frequency of forecast submissions is an important design parameter in practice. We chose weekly submissions to reflect the real-time idea of the course, with each week's forecasts responding to the most recent input data and evaluation feedback from previous weeks. While this motivation seems coherent, weekly submissions were quite challenging to some students, and we conjecture that the weekly format emphasizes forecasting workflow (i.e., well-organized code and data, procedures for plausibility checks) over purely statistical aspects. Whether or not this emphasis is desirable is open to debate; it certainly seems reflective of some practical forecasting settings  \citep[see, e.g.,][]{TaylorLetham2018}. Alternative conceivable submission formats include monthly submissions with more forecasts horizons for each submission round (e.g.,\ 20 rather than five horizons). While this would reduce the workload for the participants, such a setting might be not suitable for all of the target variables (e.g., due to ensemble weather forecasts being available for horizons up to a few days only), and forecast evaluation would become more challenging due to the required aggregation over many forecast horizons. In addition, the practical predictability will be limited for longer forecast horizons and forecasts might merely reflect unconditional distributions. As an alternative to submitting specific probabilistic forecasts, participants could also be required to submit code that maps  an input vector $X$ of features to a set of quantile forecasts. This code would need to be submitted only once (or perhaps two to three times, if revisions are allowed), which would focus the workload on a few occasions. However, this approach would not only brush over practically relevant aspects (e.g., some features becoming unavailable at short notice), but might also be discouraging if there is no chance to correct suboptimal forecast model code. 

\paragraph{Duration of the course.} Our course spanned 14 weeks. While extending the course duration seems unproblematic, shortening the course seems trickier given students' initial investment in setting up a practical forecasting workflow, and the requirement of sufficient outcome data for forecast evaluation.

\paragraph{Level of difficulty.} Our course offering involved Bachelors, Masters and PhD students with a good quantitative background. To some degree, the difficulty of the course can likely be reduced by providing participants with more specific material (e.g., example code for alternative forecasting models) and more specific guidance (e.g., providing check lists and practical forecasting tips, recommending specific models). Conversely, the difficulty could be increased by providing less specific material and guidance. Overall, the diverse set of skills required for practical yet rigorous forecasting may imply a steep learning curve even for very qualified students. We recommend that instructors communicate this aspect in order to set students' expectations and avoid frustration. 

\paragraph{Parameters affecting teachers' workload.} The real-time format of the course implies that teachers may need to react to unforeseen developments at short notice. Key examples include missing or erroneous real-time data, or software issues related to the process of receiving, visualizing and evaluating forecasts. In order to limit the workload related to real-time data, we recommend to focus on an applied setting close to the teachers' expertise, and for which stable data sources are available. In terms of software maintenance, some core functionalities (e.g., providing basic feedback on forecast performance) also seem achievable by simple means (e.g., sending an email newsletter to course participants). 

\section*{Code and data availability}

The repository containing students' forecasts and all codes used to run the challenge is available at \hyperref[tab:urls]{URL1}. Codes to reproduce the tables and figures from this paper are available at \hyperref[tab:urls]{URL2}. This repository also contains a simplified version of the interactive visualization dashboard created for the challenge, which can be used as a starting point for similar efforts.

\section*{Acknowledgements}

We thank our students for investing considerable time and effort into this novel course format. Furthermore, we acknowledge a student award generously provided by the International Institute of Forecasters (\hyperref[tab:urls]{URL15}). Sebastian Lerch gratefully acknowledges support by the Vector Stiftung through the Young Investigator Group ``Artificial Intelligence for Probabilistic Weather Forecasting.'' Johannes Bracher acknowledges support by the Helmholtz Foundation via the SIMCARD Data Science Pilot Project. Marco Wurth and Benedikt Schulz provided excellent assistance with the weather data. Finally, we thank two anonymous reviewers for helpful comments on an earlier version of the article.

\bibliographystyle{apalike}
\bibliography{ptsfc}

\newpage

\begin{appendix}

\section{Web resources}

\begin{table}[!htbp]
\caption{Listing of web resources mentioned in the text. Last accessed on October 28, 2022. \label{tab:urls}\vspace{.4cm}}
	\begin{tabular}{lp{14cm}}
		\toprule
		Reference & URL \\	\midrule
		URL1 & \url{https://git.scc.kit.edu/ng3223/ptsfc_results} \\
		URL2 & \url{https://github.com/FK83/ptsfc_replication} \\
		URL3 &  \url{https://jobrac.shinyapps.io/ptsfc_viz/}	\\
		URL4 & \url{https://de.finance.yahoo.com/} \\ 
		URL5 & \url{https://www.dwd.de/DE/wetter/wetterundklima_vorort/berlin-brandenburg/berlin_tempelhof/_node.html} \\
		URL6 &  \url{https://www.dwd.de/EN/ourservices/nwp_forecast_data/nwp_forecast_data.html}\\
		URL7 & \url{https://en.wikipedia.org/wiki/List_of_Discworld_characters} \\
		URL8 & \url{https://en.wikipedia.org/wiki/List_of_Brooklyn_Nine-Nine_characters} \\
		URL9 & \url{https://en.wikipedia.org/wiki/List_of_Friends_and_Joey_characters} \\
		URL10 & \url{https://en.wikipedia.org/wiki/List_of_Game_of_Thrones_characters} \\
		URL11 & \url{https://en.wikipedia.org/wiki/List_of_Star_Wars_characters} \\
		URL12 &  \url{https://git.scc.kit.edu/ng3223/ptsfc_results/-/tree/main/ptsfc_viz} \\
		URL13 & \url{https://image-net.org/challenges/LSVRC/} \\
		URL14 & \url{https://s2s-ai-challenge.github.io/} \\ 
		URL15 & \url{https://forecasters.org/programs/research-awards/students/} \\ \bottomrule
	\end{tabular}
	
\end{table}

\newpage

\section{Schedule of preparatory video lectures}
\label{sec:appendix_schedule}

In the following, we describe the content of the preparatory video lectures that we made available to students at the beginning of the course (see Section \ref{sec:struct}). 

\begin{enumerate}
	\item Introduction (37 minutes)
	\begin{itemize}
		\item Aims and contents of the course
		\item Course logistics: Time schedule of weekly online meetings and forecast submissions; submission rules; grading
	\end{itemize} 
	\item General content on methodology
	\begin{enumerate}
	\item Regression models (120 minutes)
	\begin{itemize}
	\item Linear regression and extensions
	\item Model complexity, overfitting and regularization
	\item Model validation
	\item Distributional regression
	\item Trees, random forests and boosting
	\item Neural networks (for probabilistic forecasting)
	\item Combining probabilistic forecasts
\end{itemize}
\item Time series analysis (51 minutes)
\begin{itemize}
\item Trend and seasonality
\item Stationarity, autocovariance and autocorrelation
\item AR(1) process and extensions
\item GARCH models
\end{itemize}
\item Forecast evaluation (92 minutes)
\begin{itemize}
\item Consistent scoring functions and proper scoring rules for different types of forecasts: Mean; probability of binary outcome; quantiles and intervals; full distributions
\item Assessing forecast calibration: Mincer-Zarnowitz regression and PIT histogram
\item Decision theory: Consistent scoring function, Bayes acts, and elicitable functionals
\end{itemize}
\item R resources (12 minutes)
\begin{itemize}
	\item Setup and editors
	\item Important packages
\end{itemize}
\item Python resources (28 minutes)
\begin{itemize}
	\item Setup and editors
	\item Important packages and jupyter notebook
\end{itemize}
\end{enumerate}
\item Course-specific content
\begin{enumerate}
	\item Temperature and wind data (106 minutes)
	\begin{itemize}
		\item Sources of weather forecasts
		\item Numerical weather and ensemble prediction
		\item Post-processing ensemble forecasts
		\item Explanation of weather variables
		\item Benchmark models and possible extensions
	\end{itemize}
	\item DAX data (72 minutes)
	\begin{itemize}
		\item Prices and (logarithmic) returns
		\item Definition of prediction task
		\item Benchmark model and possible extensions
		\item Quantile regression
	\end{itemize}
	\item Real-time forecasting workflow (24 minutes)
	\begin{itemize}
	\item Hindsight bias and retrospective forecasting pitfalls
	\item Lessons from real-time forecasting of COVID-19
	\item Practical recommendations
	\end{itemize}
\item Forecast submission and evaluation in the course (29 minutes)
\begin{itemize}
	\item Submission rules
	\item Evaluation via quantile score and computation of overall ranking
\end{itemize}
\end{enumerate}
\end{enumerate}

In terms of additional resources on the methodological topics, the textbooks by \cite{JamesEtAl2013} and \cite{HastieEtAl2009} (on statistical machine learning), \cite{Hyndman2021} and \cite{BrockwellDavis2016} (on time series analysis), as well as the overview article by \cite{GneitingKatzfuss2014} (on forecast evaluation) may be useful for students.

\end{appendix}



\section*{Supplementary material}

# Online supplement for 'Learning to forecast: The probabilistic time series forecasting challenge'

Johannes Bracher[1,2], Nils Koster[1], Fabian Krüger[1] and Sebastian Lerch[1,2]

[1]Karlsruhe Institute of Technology
[2]Heidelberg Institute for Theoretical Studies

# 1 Additional tables

| Target | Horizon | Length (50% level) | | Length (95% level) | |
|--------|---------|-----------|----------|-----------|----------|
|        |         | Benchmark | Ensemble | Benchmark | Ensemble |
| DAX    | 1 day   | 1.1 | 1.3 | 5.1  | 4.4  |
| DAX    | 2 day   | 1.7 | 1.8 | 7.0  | 6.0  |
| DAX    | 5 day   | 2.0 | 2.2 | 8.7  | 7.5  |
| DAX    | 6 day   | 2.4 | 2.4 | 9.8  | 8.3  |
| DAX    | 7 day   | 2.6 | 2.6 | 10.9 | 9.1  |
| Temp.  | 36 hour | 1.0 | 1.7 | 2.6  | 5.0  |
| Temp.  | 48 hour | 1.1 | 1.7 | 2.8  | 4.9  |
| Temp.  | 60 hour | 1.4 | 2.0 | 3.3  | 5.8  |
| Temp.  | 72 hour | 1.7 | 2.0 | 4.8  | 5.9  |
| Temp.  | 84 hour | 2.3 | 2.4 | 5.1  | 6.9  |
| Wind   | 36 hour | 2.7 | 4.5 | 7.2  | 13.2 |
| Wind   | 48 hour | 2.8 | 4.9 | 7.8  | 14.4 |
| Wind   | 60 hour | 4.0 | 5.3 | 9.2  | 15.1 |
| Wind   | 72 hour | 3.4 | 5.1 | 8.8  | 14.8 |
| Wind   | 84 hour | 4.2 | 5.4 | 10.9 | 15.8 |

Table 1: Length of prediction intervals, on average over the sample period.

# 2 Exploratory analysis of ranking methodology

In Sections 2.1 and 2.2, we explore the informativeness and robustness of the ranking along two dimensions.

## 2.1 Stability

Here we analyze the stability of the ranking. In particular, if the ranking merely reflected small or unsystematic differences in forecast performance, or would be driven by a small number of forecast dates, it would not serve its purposes well. In order to assess the ranking's stability, we resample many possible sample paths, each of which represents one hypothetical course of history like the one

| Forecaster (Alias) | Share of Wins (in %) | Mean | Rank Quantiles 90% | 99% | Actual rank |
|---|---|---|---|---|---|
| Yoda | 48.5 | 1.8 | 3 | 5 | 1 |
| PhoebeBuffay | 31.8 | 2.3 | 4 | 6 | 2 |
| HotPie | 10.2 | 3.2 | 5 | 7 | 3 |
| Shaggydog | 3.1 | 4.1 | 6 | 8 | 4 |
| KyloRen | 1.2 | 5.1 | 7 | 9 | 5 |

Table 2: Summary of forecaster ranks obtained across 10 000 simulated sample periods, generated as described in the text. Table focuses on five best forecasters out of 21 participants.

we observed in reality. We then compare the rankings obtained across many possible histories. More specifically, we proceed as follows: In each Monte Carlo iteration, we

- draw 14 DAX forecast dates (independent draws from 14 actual forecast dates, *with* replacement). Compute DAX score rankings based on the empirical scores for these dates.

- draw 13 weather forecast dates (independent draws from 13 actual forecast dates, *with* replacement). Compute temperature and wind score rankings based on the empirical scores for these dates. We use the same dates for temperature and wind in order to account for dependence of these variables.

- construct an overall forecast ranking from the rankings in the present iteration.

This simple simulation design can be viewed as investigating the robustness of the forecast ranking across individual weeks of the empirical sample period. Table 2 summarizes the simulation results for the five best forecasters from the actual (empirical) ranking. The best-performing forecaster in the empirical sample (alias 'Yoda') also performs best across the simulations, in terms of highest share of wins, smallest mean rank, and smallest rank quantiles. Furthermore, one of either 'Yoda' or 'Phoebe Buffay' wins in more than 80% of the simulated time paths. In addition to the results from Table 2, we find that ten out of 21 forecasters never win across the 10 000 Monte Carlo iterations. We view these overall results as reassuring evidence on the ranking's signal-to-noise ratio.

## 2.2 Comparison to alternative ranking variants

As discussed in Section 2.3 of the main paper, the ranking used in the course sums up a forecaster's ranks across 15 sub-categories (three target variables times five horizons). The forecaster's final rank is then determined by the resulting rank sum. This pragmatic approach is motivated by the fact that the scores for the sub-categories are on different scales. For example, one would generally expect higher scores at longer forecast horizons, due to less predictability. Here we compare this original ranking method ('Version 1') to two alternatives:

- Version 2: Compute a forecaster's skill score for each of the 15 sub-categories. Determine final rank from the sum of skill scores.

- Version 3: Determine a forecaster's final rank from the sum of raw scores across all forecast cases.

Version 2 is similar in spirit to Version 1, in that it transforms the scores (by computing skill scores) to make them comparable across sub-categories. By contrast, Version 3 does not account for possible scale differences. While this approach may be viewed as a drawback from a pragmatic perspective, its theoretical properties (in terms of being proper) are clearer and more attractive than for Versions 1 and 2.

Figure 1 summarizes the comparison. Naturally, the three rankings vary to some degree; for example, the second-best forecaster according to Version 1 ranks first according to Version 3 and

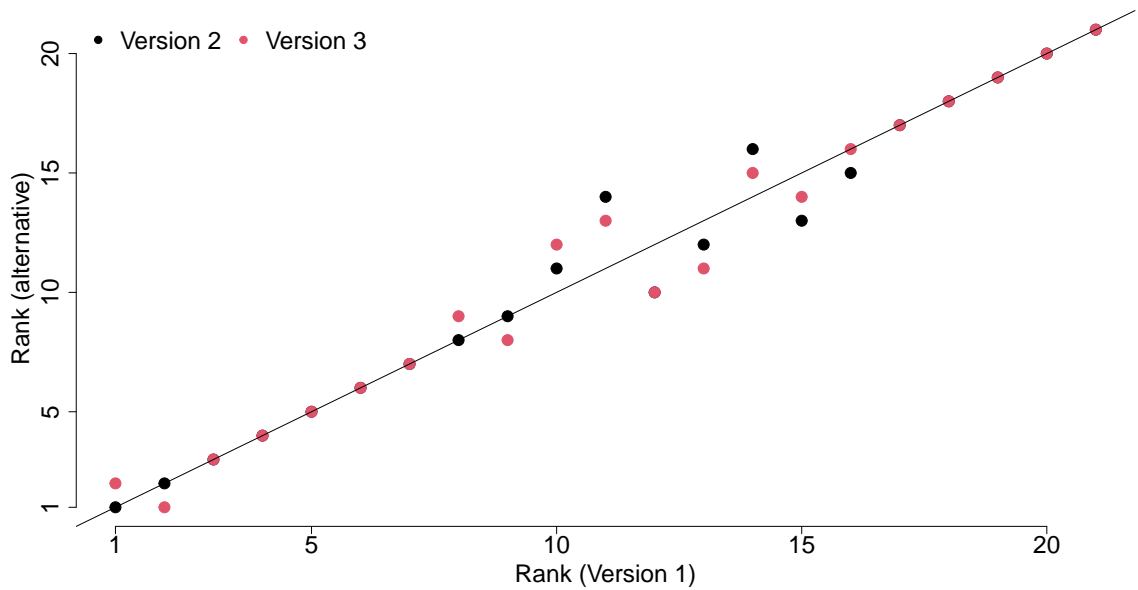

Figure 1: Comparison of forecaster ranks across three different methods. See text for details.

second according to Version 2. However, these variations seem minor overall, and the correlation between any pair of rankings is at least 98.4%.



\end{document}